# Boosting quantum yields in 2D semiconductors via proximal metal plates

*Yongjun Lee[1], Anshuman Kumar[2], Johnathas D'arf Severo Forte[3], Andrey Chaves[3,4], Shrawan Roy[1], Takashi Taniguchi[5], Kenji Watanabe[6], Alexey Chernikov[7], Joon I. Jang[8], Tony Low[9,\*] and Jeongyong Kim[1,\*]*

[1]Department of Energy Science, Sungkyunkwan University, Suwon 16419, Republic of Korea

[2]Physics Department, Indian Institute of Technology Bombay, Mumbai 400076, India

[3]Departamento de Física, Universidade Federal do Ceará, Campus do Pici, 60455-900 Fortaleza, Ceará, Brazil

[4]Department of Physics, University of Antwerp, Groenenborgerlaan 171, B-2020 Antwerpen, Belgium

[5]International Center for Materials Nanoarchitectonics, National Institute for Materials Science, 1-1 Namiki, Tsukuba 305-0044, Japan

[6]Research Center for Functional Materials, National Institute for Materials Science, 1-1 Namiki, Tsukuba 305-0044, Japan

[7]Department of Physics, University of Regensburg, Regensburg, D-93040, Germany

[8]Department of Physics, Sogang University, Seoul 04107, Republic of Korea

[9]Department of Electrical & Computer Engineering, University of Minnesota, Minneapolis, MN 55455, USA

AUTHOR INFORMATION

**Corresponding Authors**

\*T.L (tlow@umn.edu), J. K (j.kim@skku.edu)

**Abstract**

Monolayer transition metal dichalcogenides (1L-TMDs) have tremendous potential as atomically thin, direct bandgap semiconductors that can be used as convenient building blocks for quantum photonic devices. However, the short exciton lifetime due to the defect traps and the strong exciton-exciton interaction in TMDs has significantly limited the efficiency of exciton emission from this class of materials. Here, we show that exciton-exciton dipolar field interaction in 1L-WS$_2$ can be effectively screened using an ultra-flat Au film substrate separated by multilayers of hexagonal boron nitride. Under this geometry, dipolar exciton-exciton interaction becomes quadrupole-quadrupole interaction because of effective image dipoles formed inside the metal. The suppressed exciton-exciton interaction leads to a significantly improved quantum yield by an order of magnitude, which is also accompanied by a reduction in the exciton-exciton annihilation (EEA) rate, as confirmed by time-resolved optical measurements. A semiclassical model accounting for the screening of the dipole-dipole interaction qualitatively captures the dependence of EEA on exciton densities. Our results suggest that fundamental EEA processes in the TMD can be engineered through proximal metallic screening, which represents a practical approach towards high-efficiency 2D light emitters.

**Introduction**

Transition metal dichalcogenides (TMDs) are layered materials where the forces between its constituent layers are mediated by weak van der Waals interaction. In particular, the 1H phase of TMDs monolayers with $MX_2$ (M = Mo, W and X = S, Se) stoichiometry has a direct bandgap at the high symmetry K (K') points of the Brillouin zone[1,2]. Due to strong two-dimensional (2D) confinement, Coulomb bound electron-hole pairs, commonly known as excitons, can be formed upon optical excitation within the light cone at the K (K') points, where they would subsequently undergo radiative recombination to yield photoluminescence (PL)[3]. However, the quantum yield (QYs) of monolayer (1L)-$MX_2$ is fundamentally limited by various nonradiative decay pathways. This includes the conversion of bright excitonic states into spin- and valley-forbidden states (dark states) [3,4]. Numerous efforts have been recently made to improve the QY by investigating and/or even controlling the limiting mechanisms. It was found that as-exfoliated semiconducting TMDs, such as 1L-$WS_2$ and 1L-$MoS_2$, are inherently n-type, which have dense excess free carriers and thus promote the formation of trions and their nonradiative recombination[5-7]. In addition, the presence of a large number of defects in 1L-TMDs (especially chalcogen vacancies, ~$10^{12}$ cm$^{-2}$) can serve as nonradiative recombination centers leading to low QYs [8-10]. Various methods such as defect passivation or charge transfer were proposed to improve the QY [7,10-14]. For example, a near-unity QY in sulfur-based 1L-$MS_2$ has been reported through passivation of defect-mediated nonradiative recombination by super-acid treatment [12] and atomic healing of sulfur vacancies were also observed by scanning transmission electron microscopy [10]. Current understanding is that the compensation of the n-doping, such as via chemical treatment or by electrostatic doping, restores the predominance of neutral excitons, and therefore, enhance the PL efficiency of the 1L-TMDs [12-14].

However, even in the ideal pristine 2D $MX_2$, the QY can still be quenched when the exciton-exciton dipolar interaction dominates to cause strong exciton Auger recombination, so-called exciton-exciton annihilation (EEA) [15-19]. This density-dependent fundamental process occurs at high exciton densities under strong light illumination. Most notably, experimentally reported EEA rate constants revealed a very large variation across different device configurations with values ranging from a few $10^{-1}$ $cm^2/s$ to $10^{-3}$ $cm^2/s$ for 1L-$WS_2$. These observations are consistent with the sensitivity of the exciton-exciton Coulomb interaction due to its surrounding dielectric environment [20]. These reported EEA rate constants are also orders of magnitude higher than those typically observed in semiconducting quantum-well systems [21]. Since electroluminescence (EL) in 2D TMDs arises as a consequence of radiative excitonic transition [22], EEA practically determines the quantum efficiency of a light-emitting device at the exciton density under typical brightness [23,24]. Without question, EEA process represents one of the key technological bottlenecks for achieving energy-efficient and bright EL devices even in otherwise pure, high-quality samples [8]. Despite experimental reports that the use of high-index substrates [25] or hBN encapsulation [26,27] resulted in largely reduced EEA rates in 1L-TMDs accompanied by the relative increase of PL, its physical origin was not well understood. Fundamental understanding of EEA, and strategies in alleviating EEA and the accompanying QY drooping at high exciton densities are of utmost importance to high efficiency light emitting devices.

As depicted in Fig. 1a, the dipolar interaction between excitons and its Coulombic fields can be strongly influenced by the presence of a proximal metallic plate because the image dipoles in the metal renders exciton-exciton interaction essentially quadrupolar. In this work, we provide the first systematic study of EEA and QY enhancement in this device geometry using ultra-flat Au substrates with atomically controlled hBN spacers. Our simple approach of controlling the exciton interaction via proximal metal plates provides an efficient and feasible route towards practical applications using highly efficient light-emitting devices derived from 1L-TMDs.

**Results and Discussion**

Fig. 1a illustrates a schematic of the sample configuration, with 1L-WS$_2$ prepared on an ultra-flat Au surface separated by a multilayered hBN spacer to serve as insulating barrier, compared with 1L-WS$_2$ directly on SiO$_2$. Coulombic dipolar fields between two excitons would be substantially suppressed due to metallic screening by the Au substrate; namely, each image dipole aligned opposite to the real dipole makes the exciton quadrupolar, rendering mutual interaction as quadrupole-quadrupole interaction. Our goal is to investigate the anticipated reduction in Auger scattering between excitons, or EEA. In Figs. 1b and 1c, the atomic force microscopy (AFM) image and the PL intensity map of the actual sample are shown, respectively, where the thickness of the hBN layer on the Au surface was measured to vary in the range of 7–36 nm. The PL spectra of 1L-WS$_2$ obtained on the SiO$_2$/Si substrate and on Au at several hBN thicknesses are displayed in Fig. 1d. As shown in Figs. 1c and 1d, we observed that the PL intensity is highly dependent on the hBN thickness, where its PL intensities is summarized in Fig. 1e as function of hBN thicknesses. Multiple samples were investigated, and the measured data obtained from the different 1L-WS$_2$ flakes are denoted by different symbols. We found that the PL intensities of 1L-WS$_2$ samples showed a consistent general trend, namely, a gradual increase of PL with hBN thickness up to an optimum thickness of 33 nm, followed by a rapid decrease beyond this value.

Optical interference has to be accounted for in order to understand the origin of the measured changes in the PL intensity. The detected PL is proportional to both the optical absorption and the exciton emission yield of 1L-MX$_2$, hence it is critically affected by optical interference due to the multi-layered substrate [28,29]. To properly account for the optical interference effect, we modeled the optical enhancement factor (EF) as a function of the hBN thickness on the Au substrate relative to the control configuration of air-suspended 1L-WS$_2$ using the Fresnel equation under normal incidence [28] (Details for the estimation of the EF are described in the Supplementary Information). The estimated EF is overlaid (red line) with the measured PL-intensity data in Fig. 1e. The calculated EF corroborates very well with the measured PL intensity, indicating that the optical interference effect dominates the observed PL dependence on the hBN thickness.

For direct comparison, measured the PL intensity of 1L-WS$_2$ on the SiO$_2$/Si substrate, which is denoted by the star on the left Y axis in Fig. 1e. At first glance, it seems that the relative PL intensity from 1L-WS$_2$ on the SiO$_2$/Si substrate is quite similar to that from 1L-WS$_2$ directly on Au. However, we confirmed that the actual QY of the former is five times smaller since the SiO$_2$/Si substrate gains the PL brightness via the EF by more constructive inference of multiple internal reflections, as is denoted by the star on the right Y axis in Fig. 1e. This clearly suggests that the proximity to the Au surface increased the QY of 1L-WS$_2$. As a possible explanation, the proximity to Au could have resulted in depletion of the excess electrons of 1L-WS$_2$ due to the lower work function of Au than 1L-WS$_2$ [30], and compensating hole doping has been shown to enhance the QY of n-type 1L-TMDs [11,14]. Under low photoexcitation intensities, the PL spectra of 1L-WS$_2$ are understood to be dominated by neutral exciton emissions. However, with increasing the excitation intensity, prior studies have found the emergence of the trion emission peak due to electrons generated from photoionized shallow donors [6,31]. In Figs. 1f and 1g, we show the normalized measured PL spectra observed from 1L-WS$_2$ on SiO$_2$/Si and hBN/Au substrates under various excitation intensities, respectively. With increasing the photo-excitation intensity, a low-energy shoulder in the PL spectra distinctively grew for the sample on the SiO$_2$ substrate, indicative of radiative recombination of trions. On the other hand, no additional low-energy shoulder in the PL spectra was visible for the sample on the hBN/Au substrate, which is consistent with our hypothesis of electrostatic depletion of electrons in 1L-WS$_2$ due to the higher conduction band edge of 1L-WS$_2$ (4 eV) than the Fermi level of Au (5.2 eV) [32,33].

The proximal Au surface also showed a notable impact on the lifetime of photoexcited excitons in 1L-WS$_2$. We first considered the low photo-excitation regime where EEA can be neglected. Fig. 2a shows the PL intensity map of 1L-WS$_2$ flakes prepared on the SiO$_2$/Si substrate, of which some parts are sitting on the hBN spacer layer of 27 nm thickness (highlighted by white dotted line in Fig. 2a). As shown in the PL intensity map (Fig. 2a) and the representative PL spectra obtained from 1L-WS$_2$ on SiO$_2$/Si and hBN/SiO$_2$/Si (Fig. 2b), the PL intensity from 1L-WS$_2$ on hBN/SiO$_2$/Si was lower than its counterpart on SiO$_2$/Si, whose difference can be accounted for by the optical interference effect previously discussed (See the Supplementary Information). In 2D research community, hBN is frequently used as substrate and encapsulation layers for TMDs, since it preserves the pristine quality of TMDs and has the effect of refining the spectral properties such as narrowing of the spectral linewidth [26,27,34,35]. However, the role of hBN on the PL lifetime is rarely discussed. It was recently argued that hBN encapsulation alleviates EEA but also leads to an overall reduction in the PL lifetime [26]. A separate study also reported the reduction of PL lifetime in hBN-encapsulated 1L-WS$_2$ and attributed it to enhanced exciton diffusion towards nonradiative traps [27].

We measured the PL lifetimes of 1L-WS$_2$ on hBN/SiO$_2$/Si and SiO$_2$/Si by time-resolved photoluminescence (TRPL), and their representative PL decay curves are shown in Fig. 2c. The extracted PL lifetimes were found to be 940 ± 65 ps and 143 ± 38 ps for the SiO$_2$/Si and hBN/SiO$_2$/Si substrate, respectively as shown in Fig. 2d). Due to the very low pump fluence of 2.6 nJ/cm$^2$ (corresponding to an initial exciton density of 3.6×10$^8$ cm$^{-2}$), EEA can be ignored and the PL curves indeed exhibit a simple trend of a single exponential decay. In this regime, the PL lifetime of 1L-TMDs should be mainly determined by the rate of nonradiative capture by lattice defects [9]. The reduced PL lifetime of 1L-WS$_2$ on hBN/SiO$_2$/Si is consistent with the previous studies [26,27].

Then, we examined the PL lifetimes for Au substrates. Figs. 2e and 2f show the PL decay curves and the average PL lifetimes obtained from 1L-WS$_2$ on hBN/Au, and directly on Au. The PL lifetime of 1L-WS$_2$ prepared on hBN/Au and Au exhibits average lifetime of 1008 ± 220 ps and 897 ± 89 ps, respectively, which are comparable to the values of 1L-WS$_2$ prepared on the SiO$_2$/Si substrate, as shown in Fig. 2c. We observed similar trends for all the thicknesses of hBN below 20 nm, including the case for no hBN layer on the Au substrate. To the best of our knowledge, this is the first observation of the PL lifetime increase in 1L-TMDs on hBN due to the proximal Au surface. Our observation here reconciles with the concurrent enhancement of the PL efficiency due to electron depletion discussed previously in Fig. 1f, and similarly to the PL enhancement in n-type 1L-MoS$_2$ and 1L-WS$_2$ with p-doping and defect passivation [12-14].

We now discuss the effect of the Au substrate on the exciton dynamics in 1L-WS$_2$ at high exciton densities, where exciton-exciton interaction cannot be neglected. We obtained the PL transients for different exciton densities by varying pump fluences (labeled with colors shown in the legend) for our samples with hBN thicknesses of 36, 16, and 7 nm between 1L-WS$_2$ and Au. The corresponding normalized time traces are displayed in Figs. 3b-d. The normalized PL transients of 1L-WS$_2$ on the SiO$_2$ substrate are included for comparison in Fig. 3a. As the density of photoexcited excitons in 1L-WS$_2$ increases, the PL decay time gradually decreases, and such density-dependent exciton decay is characteristics of EEA in 1L-TMDs [15-18].

While Figs. 3b-d clearly show that the PL lifetimes decrease with increasing exciton densities on hBN/Au substrates, the overall effect is weaker for thinner hBN layers. For example, the short component of the PL lifetime of 1L-WS$_2$ on the 7-nm-thick hBN on the Au surface decreased from 919 ps at $1.24\times10^8$ cm$^{-2}$ to only 376 ps at $1.44\times10^{11}$ cm$^{-2}$, which is a much smaller reduction (by about factor of 3) as compared to that of the sample on the SiO$_2$ substrate, where it decreases from 890 ps at $8\times10^8$ cm$^{-2}$ to 122 ps at $1.95\times10^{11}$ cm$^{-2}$. To quantify the density-dependent PL decay rate, we used the rate equation to fit the PL decay curves [16,18]:

$$\frac{dn_{ex}}{dt} = -\frac{n_{ex}}{\tau_0} - \gamma n_{ex}^2 \tag{1}$$

where $n_{ex}$ is the exciton density, $\tau_0$ is the measured PL lifetime at a low exciton density, and $\gamma$ is the EEA rate constant. The solution to Eq. 1 can be simplified to $n_{ex}^{-1}(t) \cong \gamma\tau_0 \exp(t/\tau_0)$ [27], where the coefficient $\gamma$ is extracted from the slope of $n_{ex}^{-1}(t)\tau_0^{-1}$ vs $\exp(t/\tau_0)$ as shown in Fig. 3e (Justification of this approach is provided in the Supplementary Information). Fig. 3f shows the plot of the estimated $\gamma$ value vs. the hBN thickness. Note that $\gamma$ of 1L-WS$_2$ on SiO$_2$ of $(1.26 \pm 0.40) \times 10^{-1}$ cm$^2$/s is consistent with previous reports on 1L-WS$_2$ [16,18,36]. Such strong exciton interaction will critically degrade the QY of 1L-TMDs at high exciton densities. The estimated $\gamma$ value for 7-nm-thick hBN on the Au surface, however, was only $(1.78 \pm 0.94) \times 10^{-2}$ cm$^2$/s, an order of magnitude lower than $\gamma$ of that prepared on the SiO$_2$ substrate. The dramatic improvement in the exciton lifetime at high exciton densities strongly suggests that the Au surface is very effective in screening the exciton-exciton interaction, rendering EEA much less effective. Peculiarly, directly on the Au surface where the metal screening should be maximized, $\gamma$ was measured to be $(5.43 \pm 0.61) \times 10^{-2}$ cm$^2$/s, which is larger than those on the hBN layers on Au. We believe that the direct contact of 1L-WS$_2$ to the Au surface may have affected the measurement of $\gamma$ through the possible charge inhomogeneity of the Au surface and the expediated exciton decay due to energy or charges transfer. (We ignore hereafter the $\gamma$ value for direct contact with Au in discussing the trend of $\gamma$ with the thickness of hBN.) Regardless, we emphasize the robust general trend of decreasing $\gamma$ with the increased proximity of 1L-WS$_2$ to the metal substrate.

To quantitatively assess the effect of metal screening on exciton emission of 1L-WS$_2$ at high exciton densities, we measured the absolute QY (a-QY), the ratio of the emitted photon number to the absorbed photon number, of 1L-WS$_2$ prepared on the Au surface with the 7-nm-thick hBN spacer as a function of generation rate (Fig. 4a) (Details for the a-QY measurement of 1L-WS$_2$ and the estimation of the generation rate from the excitation intensity are described in the Supplementary Information). The a-QY of 1L-WS$_2$ on the SiO$_2$ substrate was also measured under the same experimental conditions. The plots of the a-QY vs. the generation rate ($G$) were fitted (solid curves) with the modified rate equation in Eq. 1, by adding generation term, $G$ for continuous-wave (CW) excitation with the EEA rate constants as fitting constants (fitting process and result are discussed in detail later). The a-QYs of 1L-WS$_2$ either on hBN/Au or SiO$_2$ substrates exhibit a stiff droop, regardless of the substrate configuration, as the excitation or exciton density increases, which is typical of EEA. However, the onset of the QY droop for the two cases is quite different; more specifically, the QY droop in the hBN/Au substrate occurred at much higher excitation levels, herein found to be more than an order of magnitude higher, strongly supporting the conclusion that EEA is effectively suppressed in this structure. For example, we provide the PL spectra of 1L-WS$_2$ (adjusted with the substrate effect on absorption and emission to show the relative QYs) on hBN/Au (black) and on SiO$_2$ (orange) at $G$ of ~6.8 × 10$^{16}$ cm$^{-2}$s$^{-1}$ and ~7.5 × 10$^{20}$ cm$^{-2}$s$^{-1}$ (indicated with light blue arrows in Fig. 4a) in Figs. 4b and 4c, where 3 times and 18 times enhancements in the a-QY were observed, respectively, clearly demonstrating the role of the hBN/Au substrate in effectively screening the Coulomb interaction between excitons, thereby suppressing EEA.

In what follows, we present a theoretical model to understand the effects of the proximal metal gate on EEA and QY. From classical electrostatics, the introduction of a perfect conductor (i.e., a metal) in a system containing a dipole screens the starting dipolar field. The reason for this is that the metal effectively creates mirror charges opposite to the original ones, transforming the system into a quadrupole, which has a lower electric field [37]. We model excitons as a classical electric dipole with a charge separation distance of 10 Å, which is approximately the Bohr radius of excitons in TMDs [38,39] By solving the Poisson equation for various numbers of hBN layers on Au for a system consisting of four regions, each with a distinct dielectric constant, representing the metal (Au), multilayer hBN, single-layer WS$_2$ and vacuum, the solution of the electrostatic potential at the 1-L WS$_2$ region due to an electron is:

$$\Phi(\rho, z) = \frac{e}{4\pi\varepsilon_0} \int_0^\infty dk \frac{J_0(k\rho)}{\varepsilon(k)} \qquad (2)$$

where $J_0(k\rho)$ is the zeroth-order Bessel function of the first kind and $\varepsilon(k) = \varepsilon_{WS_2} / [A(k)e^{kz} + B(k)e^{-kz} + 1]$ is the static dielectric function in reciprocal space. $A(k)$ and $B(k)$ are coefficients to be determined through the application of the electrostatic boundary conditions at the interfaces and were calculated using a transfer matrix scheme proposed in Ref. =40. Note that these coefficients depend on the size of the multilayer h-BN region, since its variation implies on a change of the h-BN/WS$_2$ and WS$_2$/vacuum interface positions. The electrostatic interaction between the source electron and a hole in the same region (z = 0) at a distance $\rho$ from the origin is thus $V_{eh}(\rho) = e\Phi(\rho)$. To calculate the electrostatic energy between two dipoles we sum the contributions of all pairs of charges, ignoring the electron-hole energies within each dipole, since it will not be influenced by the dipole-dipole distance. The result is

$$V(\rho_{ee},\rho_{hh},\rho_{12},\rho_{21}) = e\left[\Phi(\rho_{ee}) + \Phi(\rho_{hh}) - \Phi(\rho_{12}) - \Phi(\rho_{21})\right] \tag{3}$$

where the arguments are the electron-hole, electron-electron and hole-hole distances, respectively. Note that interactions between charges with the same (opposite) sign are positive (negative) due to the repulsive (attractive) force between them.

According to Fermi's golden rule, the scattering time is inversely proportional to the scattering probability, which is given by

$$\frac{1}{\tau_{EEA}} \propto \sum_k \left|\langle \Psi_{k+q}|V|\Psi_k \rangle\right|^2 = V^2 \sum_k \left|\langle \Psi_{k+q}|\Psi_k \rangle\right|^2 \tag{4}$$

where $\tau_{EEA}$ is the average time between scattering processes, $\Psi_k$ and $\Psi_{k+q}$ represent the initial and the scattered states and $V$ is the scattering potential. If $V$ varies slowly in space relative to the wave function of the system, then it can be removed from the integral. For what comes next, it is convenient to switch from the relative coordinates of the particles to the distance between center-of-masses and relative angular orientation. Without loss of generality, we fixed the origin as the center-of-mass of the first exciton, whose dipole points towards the positive y axis, while the center of mass of the second exciton is set at a distance $d$ and oriented in-plane at an angle $\theta$ with respect to the x axis, as shown in Fig. 5a inset. Since the dipolar excitons are oriented randomly in-plane, we used an angular average of the squared electrostatic potential energy as follows:

$$\left\langle V^2(d,\theta)\right\rangle_\theta = \frac{1}{2\pi}\int_0^{2\pi} V^2(d,\theta)d\theta \tag{5}$$

$V(d,\theta)$ is plotted in Fig. 5a as a function of the relative orientation angle, $\theta$ for 7-nm-, 36-nm- and 1L- hBN layers on Au and on $SiO_2$ at a distance corresponding to the exciton density of $1 \times 10^{11}$ cm$^{-2}$ using $n_{ex} = 1/d^2$. As seen, the potential curves are sinusoidal, much like the long-range dipole interaction. The closer the metal is to the dipole-dipole system, the weaker is the averaged scattering potential. We calculated the averaged potential ($\langle V^2(d,\theta)\rangle_\theta$) as a function of exciton density as shown in Fig. 5b. It shows the overall increasing interaction energy with increasing the exciton density but the influence of proximal metal in screening this scattering potential energy is also evident. From Eq. 4, this implies that the metal effectively increases the EEA scattering time $\tau_{EEA}$. Interestingly, in Fig. 5b, the difference in interaction energy between hBN on Au and SiO$_2$ decreases with increasing the exciton density, which reflects the fact that the proximal metallic screening is less effective when the exciton-exciton separation becomes comparable with the hBN spacer thickness. Fig. 5b also shows that 1L-hBN layer on Au could provide a much stronger screening than $N = 21$ (smallest hBN thickness in our measurement), suggesting the efficacy to suppress EEA and boost the QY with our scheme by further reducing the thickness of hBN. As shown in Fig. 5c, the measured EEA rates at various hBN thicknesses on Au and at SiO$_2$ substrates via the TRPL experiment follows the same trend as $\langle V^2(d,\theta)\rangle_\theta$ calculated at exciton density of $1 \times 10^{11}$ cm$^{-2}$ with a single scaling factor.

Furthermore, we related the a-QYs measured by CW laser excitation to the EEA rate by deriving an expression for the a-QYs from the rate equation [41]:

$$QY = QY_0 \left\{ 2 \cdot \frac{\sqrt{1+G/G_0}-1}{G/G_0} \right\} \tag{6}$$

where QY$_0$ is the QY at low excitation intensity (non-EEA regime) and $G_0 = 1/4\gamma\tau_0^2$. In Fig. 5d, we showed the results of fitting the QY versus $G$ data to Eq. 6 with the EEA rates as fitting constants. The fitting curves for hBN thickness of 7 nm on Au and for SiO$_2$/Si substrate are shown in Fig. 4a. The EEA rates determined from fitting a-QY measurement obtained by CW laser excitation are somewhat higher than those determined via time-resolved measurements that used the pulsed excitation. This discrepancy was also observed previously [27] and could originate from transient reduction of the exciton density due to diffusion of exciton at the central part of laser excitation in the TRPL measurement. However, the qualitative trend of increasing EEA rates with hBN thickness are observed in both CW and pulsed excitations, as shown in Fig. 5d.

**Conclusion**

We have demonstrated that exction-exciton interaction in 1L-WS$_2$ can be effectively screened by using hBN/Au substrates, improving the QY by more than an order of magnitude at the high exciton density regime. Spectroscopic signatures of PL intensity and lifetime, and EEA rate constants, all showed the systematic dependence on the atomic thickness between 1L-WS$_2$ and Au plate, whose qualitative features can be accounted by a semi-classical model of dipolar exciton fields in proximity to a metal. Our result provides a simple scheme to regulate the fundamental process of EEA and paves the way for highly luminescent 2D light-emitting devices based on 2D-TMDs.

**Methods**

*Sample preparation*

hBNs and 1L-WS$_2$ were mechanically exfoliated onto PDMS from bulk crystals: WS$_2$ bulk crystals were purchased from HQ graphene and hBN bulk crystals were provided by the National Institute for Materials Science, Japan. The transfer method from PDMS to a target substrate was detailed elsewhere [42]. Exfoliated hBN flakes generally come in various thicknesses and were identified by an optical microscope and subsequently stamped onto ultra-flat Au substrates (Platypus Technologies) or a 300-nm-thick SiO$_2$ layer on a Si wafer at the substrate temperature of 100 °C. Here, ultra-flat Au substrates were used to reduce the substrate roughness which may induce strain or incidental doping to 1L-WS$_2$. Our samples were then annealed at 150 °C in a vacuum oven for 12 hours. The PL from the exfoliated 1L-WS$_2$ on PDMS was measured before the transfer. 1L-WS$_2$ samples were subsequently transferred onto the hBN/Au or hBN/SiO$_2$/Si substrates kept at 70 °C and then were annealed at 70 °C in the vacuum oven for 1-2 hours. The thicknesses of hBN flakes were confirmed by atomic force microscope (XE-120, Park Systems).

*Optical measurements*

PL images and spectra were obtained with a commercial confocal microscope (Alpha-300S, WITec Instrument GmbH) equipped with a 100x objective lens (N.A.=0.9) and a frequency-doubled neodymium-doped yttrium aluminum garnet laser for 532-nm CW excitation. For TRPL measurements, the same microscope was used under pulsed excitation at 488 nm (BDL-488, Becker-Hikl GmbH), with the pulse width of 70 ps and the repetition rate of 80 MHz and an high-speed hybrid detector (HPM-100-40, Becker- Hikl GmbH) and a time-correlated single-photon counting module (TCSPC, Becker- Hikl GmbH). All measurements were conducted at room temperature.

**Acknowledgements**

J.K. and J.I.J. acknowledges the financial support by the Samsung Research Funding & Incubation Center of Samsung Electronics, under project no. SRFC-MA1802-02. Y.L. acknowledges National Research Foundation of Korea (NRF) funded by the Ministry of Education (NRF-2019R1I1A1A01062026). A.K. acknowledges funding support from the Department of Science and Technology (DST). A. Chaves and J.D.S.F. acknowledge financial support by the Brazilian Research Council (CNPq), through the PRONEX/FUNCAP and PQ programs. A. Chaves also acknowledges financial support by the Research Foundation - Flanders (FWO). K.W. and T.T. acknowledge support from the Elemental Strategy Initiative conducted by the MEXT, Japan, Grant Number JPMXP0112101001, JSPS KAKENHI Grant Number JP20H00354 and the CREST(JPMJCR15F3), JST.

**Author contributions**

T.L. and J.K. conceived and supervised the project, analyzed the data, and wrote the manuscript. Y.L. prepared the samples, performed the optical measurements, analyzed the data and wrote the manuscript; A.K., J.D.S.F., A. Chaves and T.L. performed theoretical calculation and wrote the manuscript. S.R. and Y.L. measured QY. T.T. and K.W. synthesized and discussed the role of high-quality hBNs. A. Chernikov and J.I.J. analyzed the data and wrote the manuscript.

**Competing Interests statement**

The authors declare no competing financial interests.

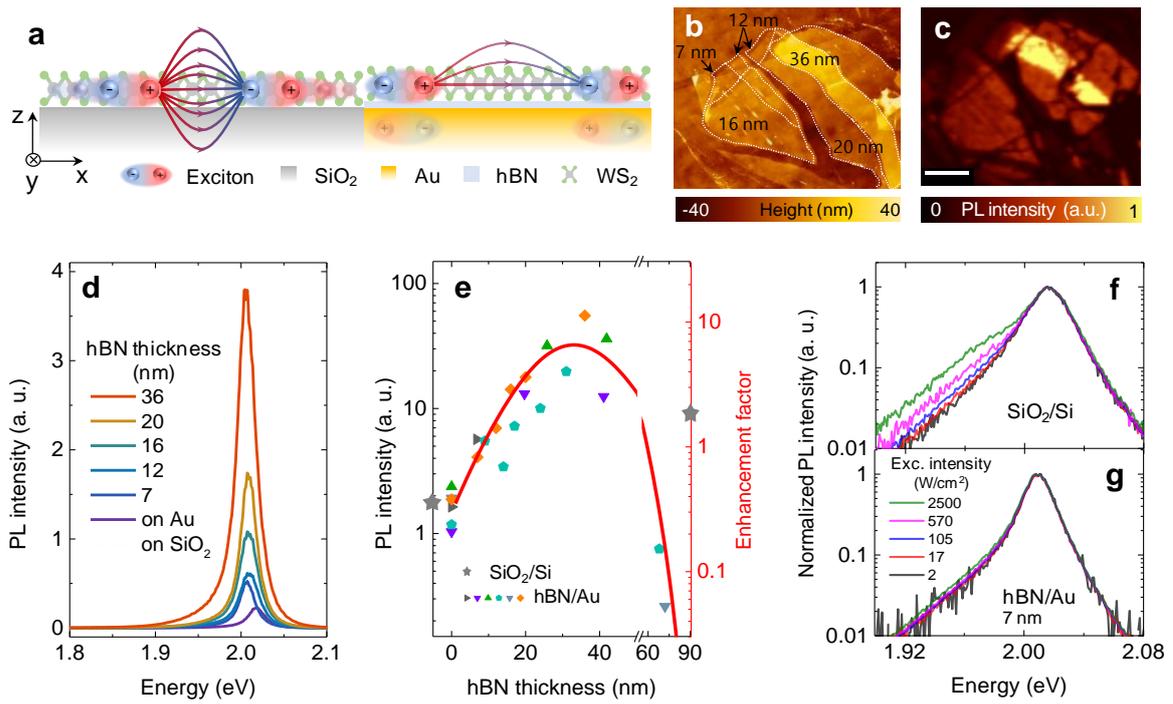

**Figure 1. hBN/Au substrate. a**. Schematic of exciton-exciton interaction suppressed by the hBN/Au substrate. Shaded arrows represent the dipolar exciton fields. Diluted field lines schematically indicate the reduced interaction by the effect of image dipoles. **b**. AFM height and **c**. PL intensity maps of 1L-WS$_2$ on the hBN/Au substrate. The scale bar in **c** corresponds to 7 μm. **d**. Measured PL spectra of 1L-WS$_2$ on various thicknesses of hBN on Au. **e**. PL intensities of six 1L-WS$_2$ samples (same color and shape of dots represent the sets of data obtained from the same 1L-WS$_2$ flakes) as a function of hBN thickness on the Au surface at the excitation intensity of 2.1 W/cm$^2$. The red curve is the calculated enhancement factor (EF) due to the optical interference of multilayer substrate relative to the suspended 1L-WS$_2$. The measured PL intensity and calculated EF of 1L-WS$_2$ on SiO$_2$/Si are indicated by the stars on the left and right Y-axes for comparison. Normalized PL spectra obtained from **f**. 1L-WS$_2$ on SiO$_2$/Si and **g**. hBN (7 nm)/Au surfaces, respectively, under various excitation (Exc.) intensities. Above 100 W/cm$^2$, a trion peak at the low-energy shoulder near 1.98 eV appeared.

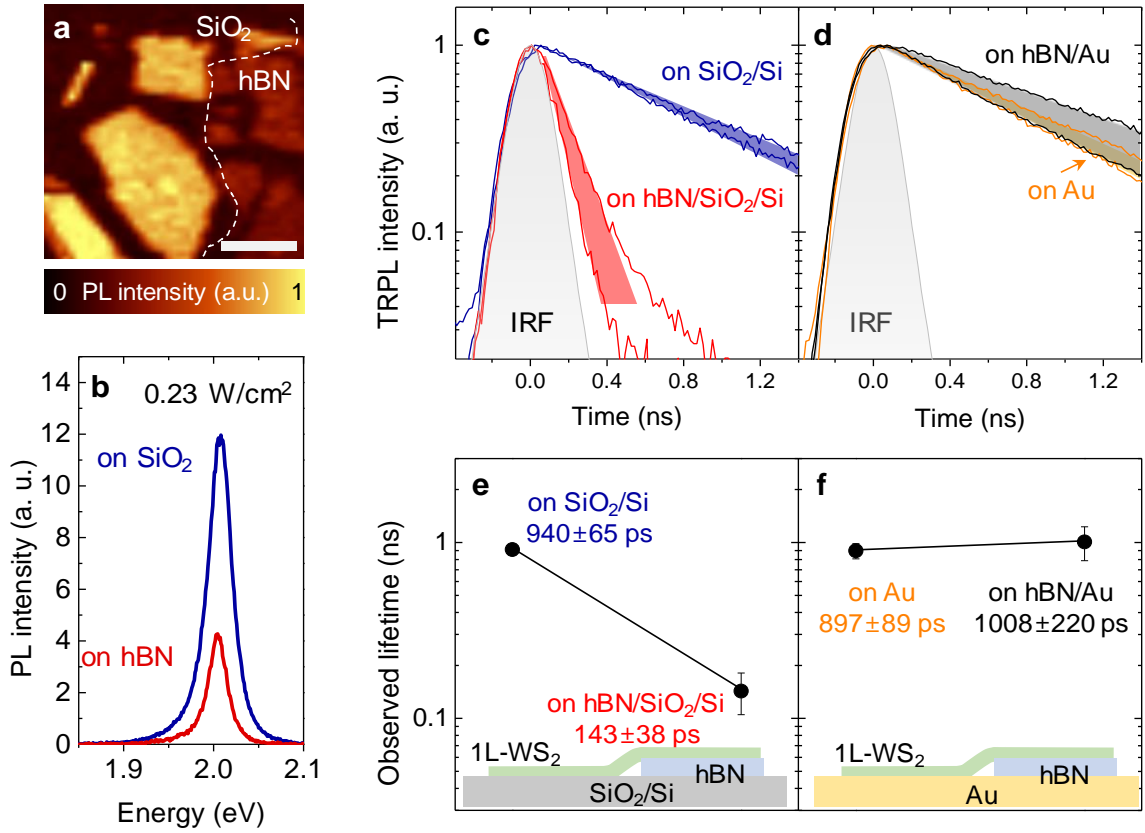

**Figure 2. Low exciton density regime**. **a**. PL intensity image and **b**. PL spectrum of 1L-WS$_2$ on SiO$_2$ (dark blue) out of the dashed region in **a** and that for 1L-WS$_2$ on hBN/SiO$_2$ (dark red) from the dashed area in **a**, respectively. The scale bar in **a** corresponds to 4 μm. **c**. Normalized TRPL data obtained at an exciton density below $5\times10^8$ cm$^{-2}$, free from EEA, from **c**. 1L-WS$_2$ on SiO$_2$ (dark blue) and hBN/SiO$_2$ (red) and **e**. 1L-WS$_2$ on Au (orange) and hBN/Au (black), respectively. The colored shades in **c** and **e** indicate the range of the data trace, error bars observed from other samples with the same geometries. Corresponding PL lifetimes are shown with the values in **d** and **f** with the same colors of the substrate conditions in **c** and **e**. The schematic structures of the samples are displayed, in which the solid lines are the guides to the eye.

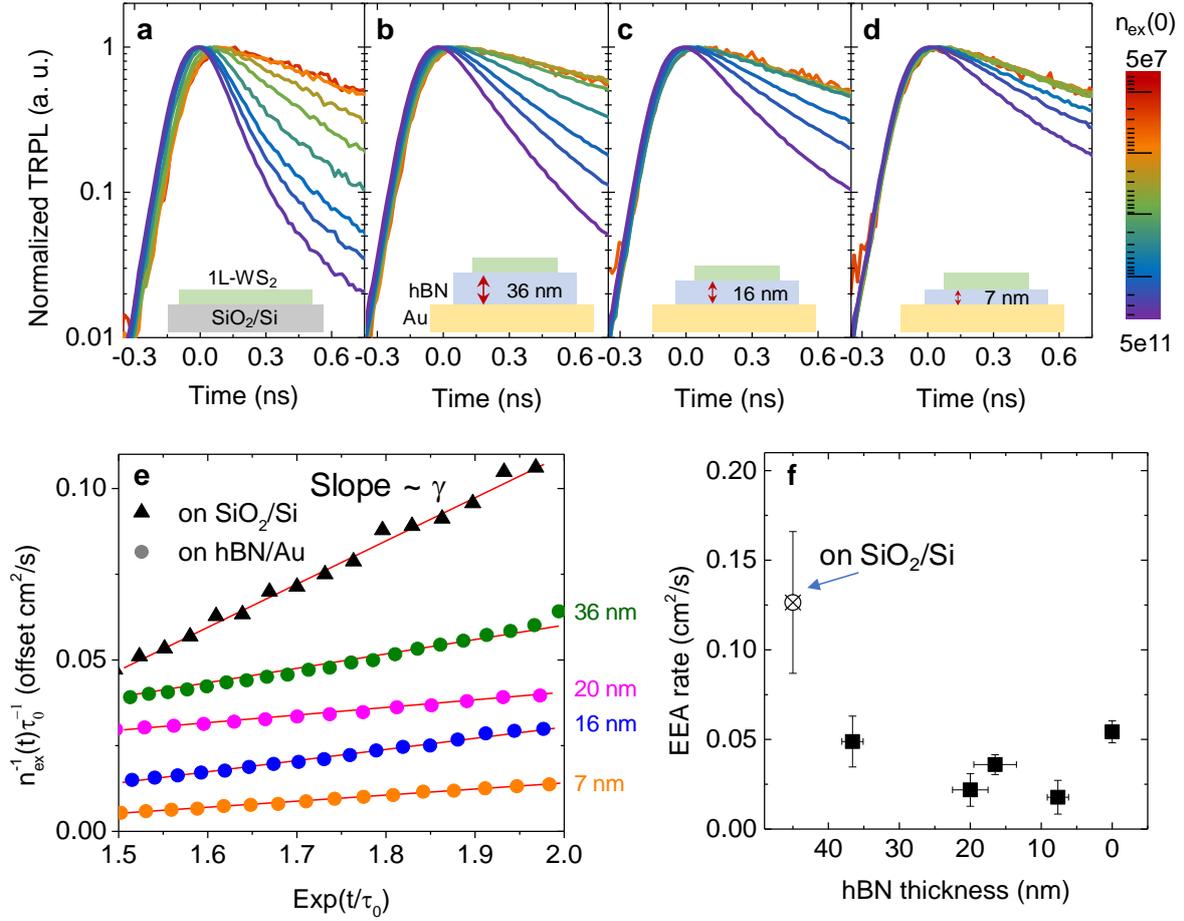

**Figure 3. Exciton dynamics vs. hBN thickness. a-d.** Series of the normalized TRPL decay curves observed from 1L-WS$_2$ on several substrate geometries under various excitation levels (initial exciton densities). The corresponding substrate geometries are shown in the insets. The color codes in the right of **d** represent the initial exciton densities. **e.** Plot of $n_{ex}^{-1}(t)\tau_0^{-1}$ vs. $\exp(t/\tau_0)$ for 1L-WS$_2$ on SiO$_2$/Si and hBN on Au for several thicknesses of hBN. The data are vertically offset for comparison. The EEA rate is extracted from the slope of linear fitting as shown by red solid lines. **f.** Estimated EEA rate as a function of hBN thickness on Au (black squares) and directly on SiO$_2$/Si (crossed circle).

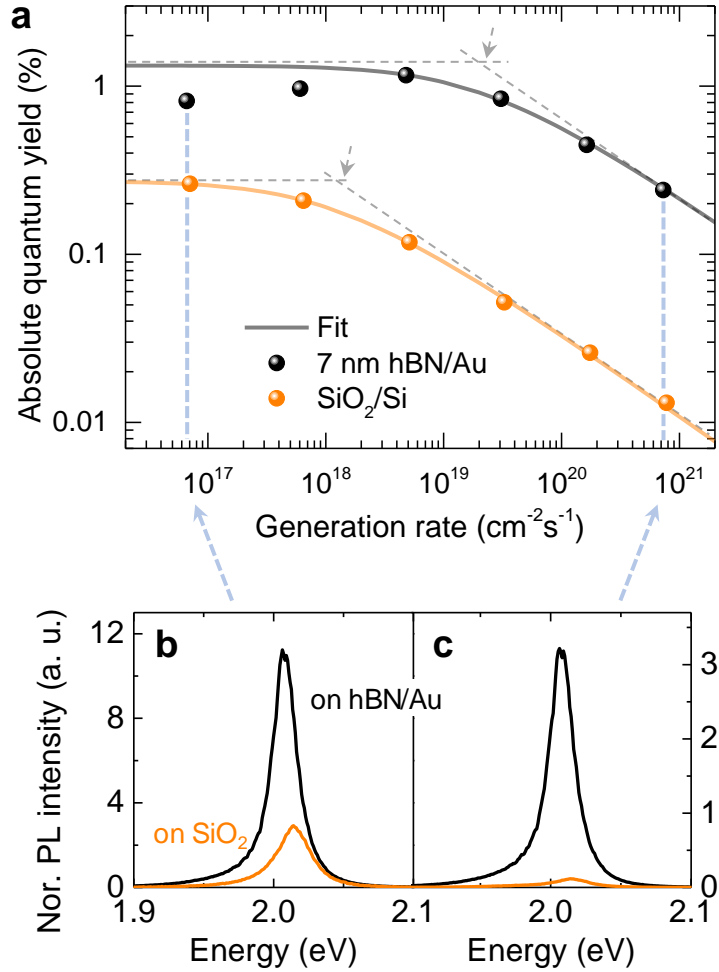

**Figure 4. Enhanced QY by suppressed EEA. a**. a-QY of 1L-WS$_2$ on hBN (7 nm)/Au (black dots) and SiO$_2$/Si (orange dots) under the generation rate ($G$) varying from $2 \times 10^{16}$ cm$^{-2}$s$^{-1}$ to $2 \times 10^{21}$ cm$^{-2}$s$^{-1}$ with the substrate interference effects properly taken into account. The solid curves with each color code are fits to the experimental QY calculated with Eq. 6. The dotted lines are guides for eyes and mark the points where the QY droop starts (grey arrows). **b** and **c**. PL spectra of 1L-WS$_2$ on hBN (7 nm)/Au (black) and SiO$_2$/Si (orange) at the $G$ indicated by light blue arrows in **a**. The PL spectra are adjusted with the substrate interference effect to show the relative a-QYs.

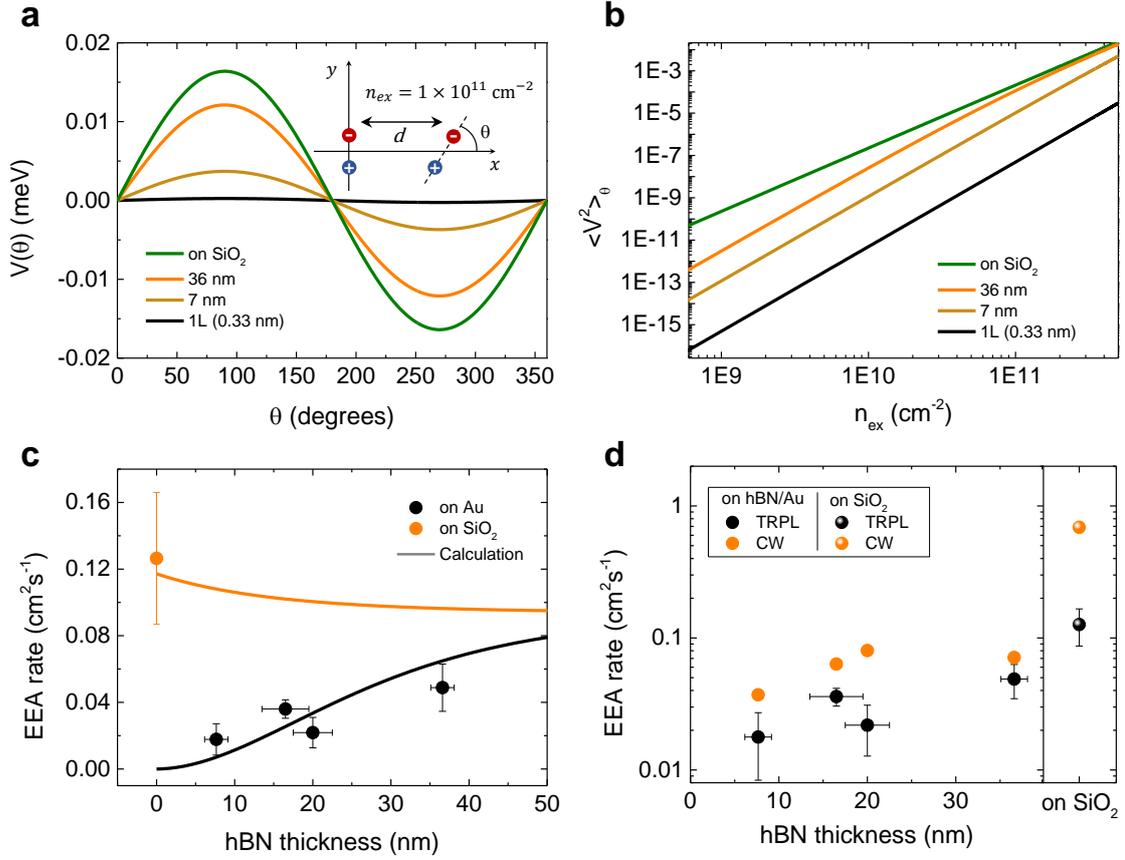

**Figure 5**. **Theoretical model. a**. The dipole-dipole interaction energy for 1L, 7nm, 36 nm thickness of hBN layers on Au and on $SiO_2$ as function of relative orientation of $\theta$ at the exciton density of $1 \times 10^{11}$ cm$^{-2}$. The effect of metal screening is different for three hBN thicknesses, which increases with the thickness of hBN. Inset: The dipole-dipole system is represented, with a dipole-dipole distance of *d* and a relative orientation of $\theta$. **b**. Log-log plot of the angle-averaged interaction energy squared versus $n_{ex}$. **c**. Measured EEA rate as a function of hBN thickness on Au (black squares from Fig. 3f) fitted with calculated $\left\langle V^2(d,\theta) \right\rangle_\theta$ (red line) from Eq. 5 at the exciton density of $1 \times 10^{11}$ cm$^{-2}$ with a proper scaling factor, showing the excellent consistency between the measured EEA rate from TRPL experiments and the calculated dipole-dipole interaction. **d**. Comparison of the EEA rates calculated from the TRPL experiments and from the CW measurements of a-QYs, showing the same trend of increasing the EEA rate with increasing the hBN thickness on Au and highest on $SiO_2$.

Supporting Information

# Boosting quantum yields in 2D semiconductors via proximal metal plates

*Yongjun Lee[1], Anshuman Kumar[2], Johnathas D'arf Severo Forte[3], Andrey Chaves[3,4], Shrawan Roy[1], Takashi Taniguchi[5], Kenji Watanabe[6], Alexey Chernikov[7], Joon I. Jang[8], Tony Low[9,\*] and Jeongyong Kim[1,\*]*

[1]Department of Energy Science, Sungkyunkwan University, Suwon 16419, Republic of Korea

[2]Physics Department, Indian Institute of Technology Bombay, Mumbai 400076, India

[3]Departamento de Física, Universidade Federal do Ceará, Campus do Pici, 60455-900 Fortaleza, Ceará, Brazil

[4]Department of Physics, University of Antwerp, Groenenborgerlaan 171, B-2020 Antwerpen, Belgium

[5]International Center for Materials Nanoarchitectonics, National Institute for Materials Science, 1-1 Namiki, Tsukuba 305-0044, Japan

[6]Research Center for Functional Materials, National Institute for Materials Science, 1-1 Namiki, Tsukuba 305-0044, Japan

[7]Department of Physics, University of Regensburg, Regensburg, D-93040, Germany

[8]Department of Physics, Sogang University, Seoul 04107, Republic of Korea

[9]Department of Electrical & Computer Engineering, University of Minnesota, Minneapolis, MN 55455, USA

AUTHOR INFORMATION

**Corresponding Authors**
  *T.L (tlow@umn.edu), J. K (j.kim@skku.edu)

**(1) Optical interference effect on the PL intensity**

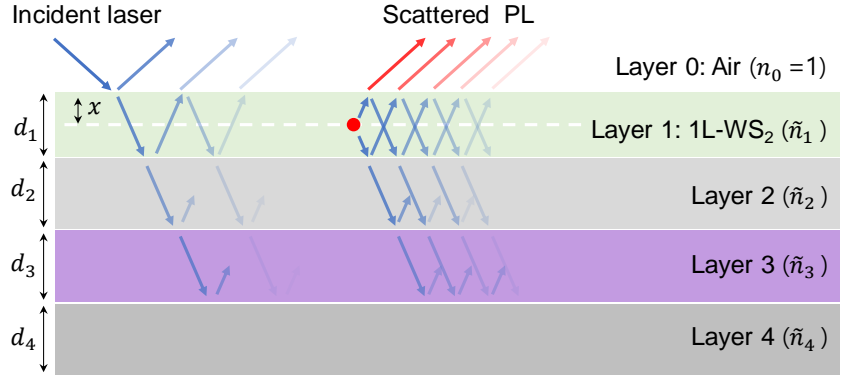

**Figure S1**. Schematic diagram of the optical interference effect in a five-layered system. (Left: absorption and right: PL emission processes) The schematic is modified from Ref. 1.

The optical interference effect can affect the observed photoluminescence (PL) intensity measured by the microscope due to multiple reflections of the light from layer interfaces [1-4]. Many researchers have considered the optical interference effect of multiple layered substrates to explain different contrast or PL intensities of two dimensional (2D) thin films such as graphene and 2D transition metal dichalcogenides (TMDs) [1-6]. Due to the multiple reflections of the light between the different dielectrics, there are multiple chances to absorb the excitation light and scatter the PL from monolayer (1L)-$WS_2$. The net absorption of excitation light and scattering of PL from 1L-$WS_2$ in the four-layered system can be expressed as [1]

$$F_{abs} = t_1 \frac{[1 + r_2 r_3 e^{-2i\beta_2}]e^{-i\beta_x} + [r_2 + r_3 e^{-2i\beta_2}]e^{-i(2\beta_1 - \beta_x)}}{1 + r_2 r_3 e^{-2i\beta_2} + (r_2 + r_3 e^{-2i\beta_2})r_1 e^{-2i\beta_1}} \quad \text{(S1)}$$

$$F_{sc} = t_1' \frac{[1 + r_2 r_3 e^{-2i\beta_2}]e^{-i\beta_x} + [r_2 + r_3 e^{-2i\beta_2}]e^{-i(2\beta_1 - \beta_x)}}{1 + r_2 r_3 e^{-2i\beta_2} + (r_2 + r_3 e^{-2i\beta_2})r_1 e^{-2i\beta_1}} \quad \text{(S2)}$$

where $t_1 = 2n_0/(\tilde{n}_1 + n_0)$, $t_1' = 2\tilde{n}_1/(\tilde{n}_1 + n_0)$, $r_n = (\tilde{n}_{n-1} - \tilde{n}_n)/(\tilde{n}_{n-1} + \tilde{n}_n)$ are the Fresnel transmission and reflection coefficients for the interfaces between different layers having their refractive indices that include air ($n_0=1$) and the 2D layer $i$ ($n_i$) (by Equation editor). Here, the prime in the transmission coefficient corresponds to the internal case, i.e., transmission from the 2D layer to air. The phase differences in Eqs. S1 and S2 are defined as $\beta_x = 2\pi x \tilde{n}_1/\lambda$ and $\beta_i = 2\pi d_i \tilde{n}_i/\lambda$, where x is a point in the depth of 1L-WS$_2$ (layer 1) in which both absorption and emission occur, $d_i$ is the thickness of the layer $i$ (the thickness of 1L-WS$_2$; $d_1$ is taken to be 0.7 nm) and $\lambda$ is the wavelength of light (532 nm for excitation and 615 nm for the PL of 1L-WS$_2$). For three- or five-layered system (the case for a suspended sample or air/1L-WS$_2$/hBN/SiO$_2$/Si), $r_3$ in Eqs. S1 and S2 is 0 or should be replaced by $(r_3 + r_4 e^{-2i\beta_3})/(1 + r_3 r_4 e^{-2i\beta_3})$, respectively.

The net emission intensity of 1L-WS$_2$ on the substrate ($I_{sub}$) can be given by

$$I_{sub} = \int_0^{d_1} |F_{abs} \cdot F_{sc}|^2 \, dx \tag{S3}$$

We also calculated the net emission intensity of the suspended 1L-WS$_2$ in air ($I_{sus}$). We defined the enhancement factor (EF) as the ratio of calculated values for various substrates ($I_{sub}$) and that for the suspended 1L-WS$_2$ ($I_{sus}$), $I_{sub}/I_{sus}^2$.

The calculated EF for the quartz substrate, 300-nm-thick SiO$_2$ on the Si substrate (SiO$_2$/Si) and 27-nm-thick hBN on SiO$_2$/Si (used in Fig. 2a in the main text) are 0.396, 1.901 and 0.527, respectively. The EF as function of hBN thickness on Au is shown in Fig. 1e (red line) in the main text. A higher EF value on a typical substrate means that 1L-WS$_2$ would have a higher PL intensity on it under the same excitation conditions even if the samples have the same quantum yields (QYs). The complex refractive indices of the materials that were used for the calculations are given in Table S1.

**Table S1.** Details of refractive indices of the materials that used for the calculations.

| Materials | Wavelength (nm) | Indices (n-ik) | | References |
|---|---|---|---|---|
| | | n | k | |
| WS$_2$ | 532 (excitation) | 4.9 | 0.9 | 7 |
| | 615 (emission) | 4.9 | 1.8 | |
| Quartz | 532 | 1.54 | 0 | 8 |
| | 615 | | | |
| SiO$_2$ | 532 | 1.46 | 0 | 9 |
| | 615 | 1.45 | 0 | |
| Si | 532 | 4.15 | 0.05 | 10 |
| | 615 | 3.9 | 0.02 | |
| Au | 532 | 0.55 | 2.2 | 11 |
| | 615 | 0.23 | 3.1 | |
| hBN | 532 | 1.85 | 0 | 12 |
| | 615 | | 0 | |

**(2) Estimation of the EEA rate**

The solution to Eq. 1 in the main text is given by [13,14]

$$n_{ex}(t) = \frac{n_{ex}^0}{\exp(t/\tau_0)(1 + n_{ex}^0 \gamma \tau_0) - n_{ex}^0 \gamma \tau_0} \qquad (S4)$$

where $n_{ex}^0$ is the initial exciton density and all the other parameters are defined in the main text. With some assumptions, Eq. S4 could be simplified as explained in below.

1) In the case of $n_{ex}^0 \ll (\gamma \tau_0)^{-1}$, the exciton density is not high enough to cause EEA. Therefore, the exciton transient shows linear decay with exciton lifetime $\tau_0$ and Eq. S4 can be simplified to $n_{ex}(t) = n_{ex}^0 \exp(-t/\tau_0)$.

**Table S2.** Summary of the parameters for EEA extraction.

| Substrates | | Lifetime, $\tau_0$ (ns) | EEA rate, $\gamma$ (cm²s⁻¹) | $n_{ex}^0$ (× 10¹⁰ cm⁻²) | | $(\gamma\tau_0)^{-1}$ (× 10¹⁰ cm⁻²) |
|---|---|---|---|---|---|---|
| | hBN thickness (nm) | | | Min. | Max. | |
| hBN/Au | 0 | 0.81 | 0.05426 | 3.1 | 11.1 | 2.05 |
| | 7 | 0.93 | 0.01776 | 44 | 202 | 6.03 |
| | 16 | 0.98 | 0.03599 | 72 | 329 | 2.83 |
| | 20 | 0.86 | 0.02185 | 82 | 374 | 5.55 |
| | 36 | 1.23 | 0.04885 | 67 | 307 | 1.66 |
| SiO₂ | | 0.94 | 0.12643 | 15 | 83 | 0.84 |

2) In the case of $n_{ex}^0 >> (\gamma\tau_0)^{-1}$, we need to consider the temporal resolution of the detector. For example, when the temporal bin is very short $t << (\gamma n_{ex}^0)^{-1}$ Eq. S4 can be simplified to $n_{ex}(t) = n_{ex}^0/(1 + n_{ex}^0 \gamma t)$, which is commonly used for the estimation of the EEA rate using a fast detection system having a temporal resolution below 100 ps [15,16]. However, if it is not sufficiently short $t >> (\gamma n_{ex}^0)^{-1}$ because of experimental limitation, the exciton decay is given by $n_{ex}(t) = (\gamma\tau_0)^{-1} \exp(-t/\tau_0)$ [13].

We summarized all the parameters that used for the assumption in Table S2. The series of TRPL transients and $n_{ex}^{-1}(t)\tau_0^{-1} = \gamma \exp(t/\tau_0)$ were displayed in Fig. S2. For all the cases, the initial exciton density satisfies well the condition for the assumption of $n_{ex}^0 >> (\gamma\tau_0)^{-1}$. In direct Au contact case, the initial exciton density is higher than $(\gamma\tau_0)^{-1}$ only by a factor of 1.4~5. Such a low initial exciton density of the TRPL measurement from the specimen on Au suggests that the EEA rate quoted in the main text could have been somewhat overestimated.

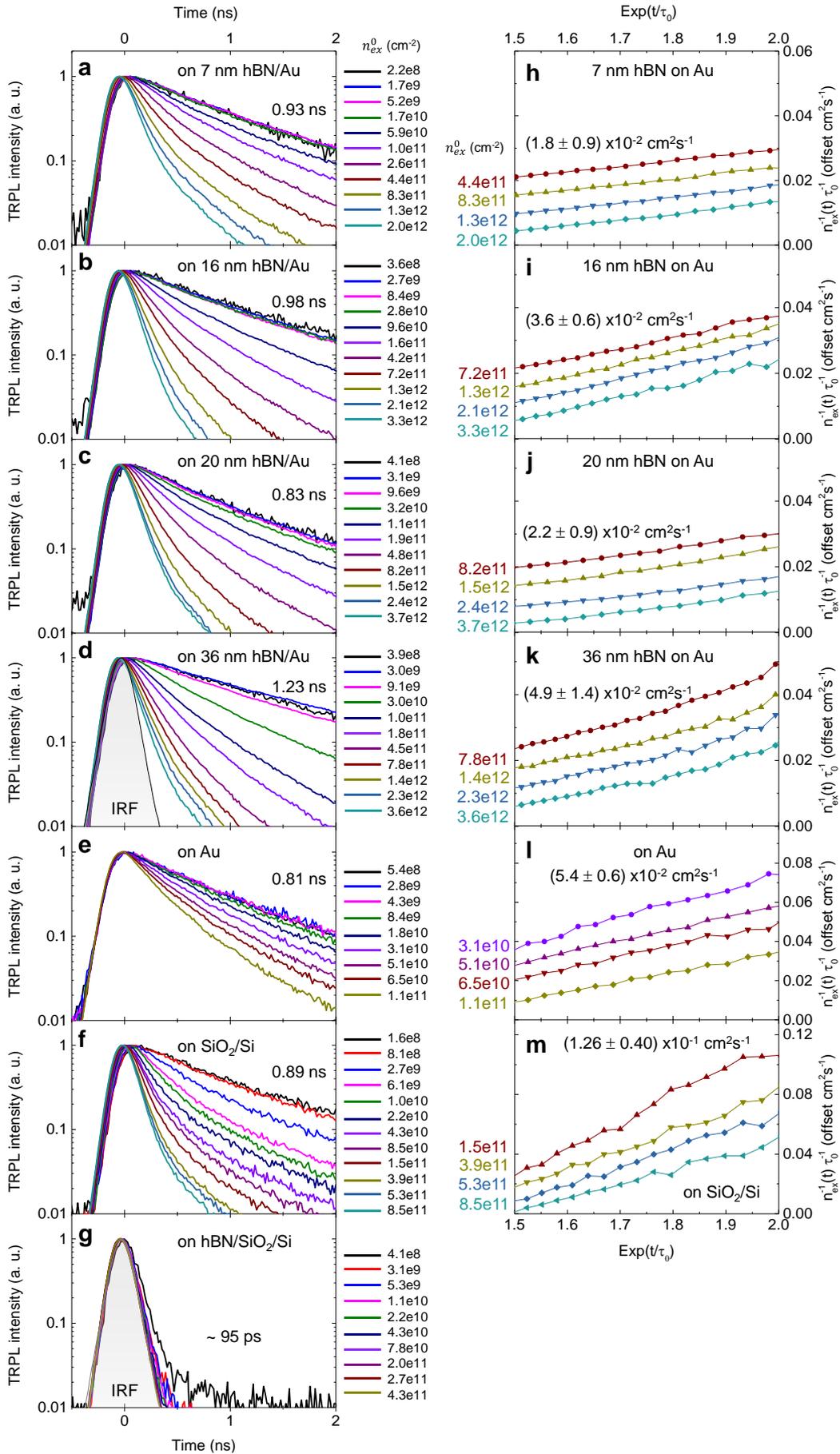

**Figure S2. TRPL transient series and EEA estimation. a-g**. TRPL decays and **h-m**. Plot of $n_{ex}^{-1}(t)\tau_0^{-1}$ vs. $\exp(t/\tau_0)$ series at each hBN thickness condition. The values of lifetimes (EEA rates) are given in **a-g** (**h-m**). The EEA rate of 1L-WS$_2$ on the hBN/SiO$_2$/Si substrate could not be extracted since the corresponding lifetime is too short to be distinguished from instrument response function (IRF). Note: The degree of absorption varies with different substrate configuration, resulting in different initial exciton densities.

**(3) Estimation of the absolute quantum yield (a-QY) using a microscope**

The integrating sphere was used for the estimation of the absolute QY (a-QY) of the samples of macroscopic sizes. In case of 1L-TMDs, however, measurements using a microscope are required due to the microscopic sizes of the samples. The details of a-QY measurement are explained in Ref. 17.

1) a-QY measurement by integrating sphere

We fabricated a poly methyl methacrylate (PMMA) thin film embedded with rhodamine 6G (R6G) as reference samples using spin coating on quartz substrates [17]. The PMMA/R6G samples are suitable for the a-QY measurement using integrating sphere (819C-SL-3.3, Newport) because it has homogeneous light emission on a centimeter scale with thickness of 80 or 300 nm (using PMMA C2 or C4, respectively). Each of the PMMA/R6G films was loaded into the integrating sphere and excited using a 532 nm laser. The scattered laser light and the PL from the film were guided to a monochromator equipped with a charge-coupled device camera through an optical fiber connected to the integrating sphere. By comparing the intensities of the absorbed 532 nm laser line and the emitted PL from the PMMA/R6G films, the a-QY values of the PMMA/R6G films were estimated. These a-QY values were double-checked using a commercial a-QY spectrometer (Quantaurus-QY, Hamamatsu Photonics) as well.

2) a-QY measurement under the microscope

To estimate the a-QY of 1L-WS$_2$, we exfoliated and transferred 1L-WS$_2$ flakes on quartz substrates with the same configuration as PMMA/R6G films. We then measured and calculated PL and absorption of the PMMA/R6G films and 1L-WS$_2$ under the same confocal microscope and detection configurations. The values of absorption of both samples were estimated using reflectance or transmittance measurements with thin film approach boundary conditions using the following relation [18]:

$$\frac{\Delta R}{R} \cong \frac{4}{n_{sub}-1}A$$
$$\frac{\Delta T}{T} \cong \frac{2}{n_{sub}+1}A$$
(S5)

By comparing absorption and PL counts of 1L-WS$_2$ and the PMMA/R6G films, we were able to estimate the a-QY of 1L-WS$_2$.